\documentclass[%
preprint,
nofootinbib,
 amsmath,amssymb,
 aps,
]{revtex4-2}

\usepackage{graphicx}
\usepackage{dcolumn}
\usepackage{bm}
\usepackage{multirow}
\usepackage{hyperref}
\usepackage{float}
\usepackage[caption = false]{subfig}

\def\pv{{\boldsymbol p}}
\def\rv{{\boldsymbol r}}
\def\ev{{\boldsymbol e}}

\def\deltav{{\boldsymbol \delta}}
\newcommand{\f}[2]{\frac{#1}{#2}}
\newcommand{\dd}{\mathrm{d}}
\newcommand{\bo}[1]{\boldsymbol #1}
\newcommand{\tH}{\theta_{\rm H}}
\newcommand{\bel}[1]{\begin{eqnarray}\label{#1}}
\newcommand{\eel}{\end{eqnarray}}
\def\bea{\begin{eqnarray}}
\def\eea{\end{eqnarray}}
\def\beq{\begin{equation}}
\def\eeq{\end{equation}}

\newcommand{\refeq}[1]{Eq.~(\ref{#1})}

\newcommand{\nn}{\nonumber}

\newcommand{\comment}[1]{}

\def\LR{\left(} 
\def\RR{\right)}
\def\LS{\left[} 
\def\RS{\right]}

\begin{document}


\title{$^3$H and $^3$He nuclei production in a combined thermal and coalescence framework for heavy-ion collisions  in
the few-GeV energy regime}

 \author{Zbigniew Drogosz}%
 \email{zbigniew.drogosz@uj.edu.pl}
\author{Wojciech Florkowski}
\email{wojciech.florkowski@uj.edu.pl}
\author{Nikodem Witkowski}%
 \email{nikodem.witkowski@student.uj.edu.pl}
\affiliation{%
 Jagiellonian University, PL-30-348 Krak\'ow, Poland
}
\author{Radoslaw Ryblewski}
\email{radoslaw.ryblewski@ifj.edu.pl}
\affiliation{
 Institute  of  Nuclear  Physics  Polish  Academy  of  Sciences,  PL-31-342  Krak\'ow,  Poland\\
}%

\date{\today}

\begin{abstract}
A thermal model describing hadron production in heavy-ion collisions in the few-GeV energy regime is combined with the idea of nucleon coalescence to make predictions for the $^3$H and $^3$He nuclei production. A realistic parametrization of the freeze-out conditions is used, which reproduces well the spectra of protons and pions. It also correctly predicts the deuteron yield that agrees with the experimental value. The predicted yields of $^3$H and $^3$He appear to be smaller by about a factor of two compared to the experimental results. The model predictions for the spectra can be compared with future experimental data. 
\end{abstract}

\maketitle


\section{Introduction}

In this work, we use a statistical hadronization model of hadron production in heavy-ion collisions in the few-GeV energy regime~\cite{Harabasz:2020sei,Harabasz:2022rdt,Kolas:2024hbg}. This approach describes well the transverse momentum and rapidity spectra of protons and pions. Moreover, it correctly reproduces the experimentally measured value of the deuteron yield~\cite{Florkowski:2023uim}, provided that, of the two options, the freeze-out corresponding to the higher temperature is selected. We consider a version of the model that incorporates spheroidal hydrodynamic expansion~\cite{Harabasz:2022rdt}, which is more realistic than the one that assumes spherical symmetry. The latter was used in the original blast-wave model by Siemens and Rasmussen~\cite{Siemens:1978pb}; see also Ref.~\cite{Florkowski:2010zz}. We note that our assumptions should be contrasted with the frequent use of boost-invariant blast-wave models at ultrarelativistic energies~\cite{Schnedermann:1993ws}, which becomes inappropriate at the lower beam energies. 

\begin{table*}[h!]
    \centering
    \begingroup
    \setlength{\tabcolsep}{13pt} 
    \renewcommand{\arraystretch}{1.2} 
    \begin{tabular}{|c|c|c|c|c|c|c|c|c|}
    \hline
    \hline
    $T$ [MeV] & $\mu_{\rm B}$ [MeV] & $\mu_{\rm I_3}$ [MeV] & $R$ [fm] & $H$ [MeV] & $\delta$ & $v_R$ & $\gamma_R$ \\
    \hline
    $70.3$ & $876$ & $-21.5$ & $6.06$ & $22.5$ & $0.4$ & $0.60$ & $1.25$ \\
    \hline
    \hline
    \end{tabular}
    \endgroup
    \caption{Thermodynamic parameters (temperature $T$, baryon chemical potential $\mu_{\rm B}$, isospin chemical potential $\mu_{\rm I_3}$), system radius $R$, Hubble expansion parameter $H$, longitudinal eccentricity $\delta$, radial flow $v_R$, and Lorentz gamma factor at the system's boundary ($r=R$) $\gamma_R$, extracted from experimental data.}
    \label{tab:params}
\end{table*}

\bigskip
Our analysis is based on the data collected by the HADES Collaboration for Au-Au collisions at the beam energy $\sqrt{s_{\rm NN}}=$~2.4 GeV and the centrality class of 10\%~\cite{Szala:2019,Szala2019a,HADES:2020ver}. The fits to the particle abundances made by us and other groups~\cite{Motornenko:2021nds} suggest two different sets of possible freeze-out thermodynamic parameters. In particular, they yield two different values of the freeze-out temperature: $T=$~49.6~MeV vs $T=$~70.3~MeV. As the analysis of deuteron production favors the case with the higher temperature \cite{Florkowski:2023uim}, here we always assume that $T=$~70.3~MeV. In this case, the system size $R$ is rather small, approximately 6~fm, hence, the probability of nuclei formation by a nucleon triplet is relatively large (compared to the low-temperature case where $R \approx 16$~fm). The thermodynamic parameters used in our calculations (temperature, baryon, and isospin chemical potentials) are given in~Table~\ref{tab:params}. The thermodynamic parameters that control the production of strangeness are not displayed here, as they are irrelevant to the present analysis.

In the collisions of gold nuclei considered, one detects on average per event about 28.7 nuclei of $^2$H, 8.7 nuclei of $^3$H, and 4.6 nuclei of $^3$He~\cite{Szala:2019,Szala2019a}. \footnote{We note that the nucleus of $^3$H is also called triton, t, and the nucleus of $^3$He is also called helion, h. In this work, for simplicity, we instead often use the symbols $^3$H and $^3$He by themselves to denote the nuclei.} Hence, about 46.5 protons are found in the bound states. The remaining average number of directly produced protons is 77.6. In the framework defined in Refs.~\cite{Harabasz:2020sei,Harabasz:2022rdt} we assume that all protons eventually detected in bound states and those measured as free particles originally form a thermal system. Thus, while comparing the predictions of our model with the data, the final results for the proton spectra are rescaled by the factor 77.6/(77.6+46.5). This implicitly assumes the existence of a certain physical mechanism that combines the remaining protons with neutrons into the light nuclei mentioned above. In the present work, we assume that this process can be interpreted as coalescence and use the standard expressions of the coalescence framework (see, for example, Refs.~\cite{Mrowczynski:2016xqm,Mahlein:2023fmx}) to predict the light nuclei spectra. Continuing the approach from Ref.~\cite{Florkowski:2023uim}, we do not use Gaussian \textit{ad hoc} parametrizations of the thermal source but rather consider initial particle production as independent thermal production within a sphere of the radius $R$ (which is consistent with thermal model assumptions adopted in Refs.~\cite{Harabasz:2020sei,Harabasz:2022rdt}). Our present work does not include contributions from decaying resonances. We have verified using THERMINATOR~\cite{Kisiel:2005hn,Chojnacki:2011hb} that for the low-temperature case they are negligible, while for the high-temperature case only the Delta resonance plays a noticeable, though very small, role.

The topic of producing light nuclei has recently attracted a lot of interest (see, for example, Refs.~\cite{Sombun:2018yqh,Blum:2019suo,Mrowczynski:2019yrr,Bellini:2020cbj,Kachelriess:2020amp,Kozhevnikova:2020bdb,Zhao:2020irc,Hillmann:2021zgj,ALICE:2021mfm,Sochorova:2021lal,Zhao:2021dka,Sharma:2022tih,STAR:2023uxk}). In relativistic heavy-ion collisions, the production of such systems is well reproduced by both the thermal approach and the coalescence model. These two approaches are usually treated as exclusive alternatives (for a short and critical review of this issue, see Ref.~\cite{Mrowczynski:2020ugu}). In this work, we present a scenario for lower energies where the initial thermal production of particles is combined with a subsequent coalescence mechanism in a consistent way, showing that the two pictures may coexist and supplement each other.

The paper has the following structure: In Sec.~\ref{sec:coal} we introduce the coalescence model and discuss the nuclei formation rate. In Sec.~\ref{sec:fm} the freeze-out model is defined. The spheroidal form of expansion is discussed in Sec.~\ref{sec:spd}. We introduce there a compact description of the distribution functions with given symmetries, which turns out to be very useful for implementing the coalescence model. Our main numerical results are presented and discussed in Sec.~\ref{sec:results}. We conclude in Sec.~\ref{sec:summary}. Throughout the paper, we use natural units: $c=\hbar=k_B=1$. The metric tensor has the signature $(+1,-1,-1,-1)$.

%
\section{Coalescence model}
\label{sec:coal}
%
\subsection{Basic concept}
%
In the coalescence model for light nuclei production, the particle spectrum is obtained as a product of proton and neutron spectra taken at a fraction of the nucleus momentum~\cite{Butler:1963pp,Schwarzschild:1963zz}. Specifically, with the proton and neutron three-momentum distributions defined by
\begin{equation}
F_{\rm p}({\pv}) = \frac{\dd N_{\rm p}}{\dd^3 p}, \quad F_{\rm n}({\pv}) = \frac{\dd N_{\rm n}}{\dd^3 p}, 
\label{eq:FpFn}
\end{equation}
the three-momentum distribution of a nucleus X with the mass number $A$ and the atomic number $Z$ is the product
\begin{equation}
\frac{\dd N_{\rm X}}{\dd^3 p_{\rm X}} = A_{\rm FR}^{\rm X} \left(\,
F_{\rm p} \left(\frac{\pv_{\rm X}}{A} \right)\right)^Z \left( F_{\rm n} \left(\frac{\pv_{\rm X}}{A} \right)\right)^{A-Z},
\label{eq:coal}
\end{equation}
where $A_{\rm FR}^{\rm X}$ is the particle formation rate coefficient discussed in more detail below, and the subscripts $\rm p$ and $\rm n$ refer to protons and neutrons, respectively.

Equation (\ref{eq:coal}) assumes the additivity of three-momenta of the  nuclei that form the considered nucleus, $\pv_{\rm X} = A \pv$. Classically, this leads to a small violation of energy conservation: for example, owing to the finite nucleus binding energy, $m_{\rm \,^3\!He} = 2809$~MeV, while $m_{\rm p}+2m_{\rm n} = (938+2\times940)$~MeV = 2818~MeV. This and other related problems connected with the conservation laws in the coalescence model are usually circumvented by reference to the quantum nature of the real coalescence process, which introduces natural uncertainty of the energies and momenta of interacting particles~\cite{Mrowczynski:2016xqm}. In any case, small differences due to the binding energy and the inequality of $m_{\rm p}$ and $m_{\rm n}$ do not affect our present analysis. Therefore, in the following, we use the approximation $m_{\rm p} \approx m_{\rm n} \approx m$, where $m$ is the mean nucleon mass, and $m_{\rm X} \approx Am$. Hence, while switching in \refeq{eq:coal} from the variable $\pv_{\rm X}$ to the particle transverse momentum $p_{\rm{\perp X}}$ and rapidity $y_{\rm X}$, we use the simple rules
\begin{equation}
p_\perp = \frac{p_{\rm{\perp X}}}{A}, \quad y = y_{\rm X}.
\end{equation}

As in both theory and experiment one usually deals with invariant momentum distributions, $E \,\dd N/(\dd^3 p)$, rather than with the forms (\ref{eq:FpFn}), it is convenient to recall that for systems with cylindrical symmetry (with respect to the beam axis $z$) studied in this work we have
\begin{equation}
\frac{\dd N}{\dd^3 p} = \frac{\dd N}{2\pi\, E\, \dd y\, p_\perp \dd p_\perp} =
\frac{\dd N}{2\pi E\, \dd y\, m_\perp \dd m_\perp},
\end{equation}
where $E$ is the on-mass-shell energy of a particle, $E=\sqrt{m^2+\pv^2}$, and $m_\perp$ is its transverse mass, $m_\perp=\sqrt{m^2+p_\perp^2}$. Therefore, from \refeq{eq:coal} we obtain 
%
\begin{eqnarray}
\frac{\dd N_{\rm X}}{E_{\rm X}\, \dd y\, m_{\rm{\perp X}} \dd m_{\rm{\perp X}}} = \frac{A_{\rm FR}^X}{\left(2\pi\right)^{A-1}}
\left(\frac{\dd N_{\rm p}}{E\, \dd y\, m_{\perp} \dd m_{\perp}}\right)^{Z}\left(
\frac{\dd N_{\rm n}}{E\, \dd y\, m_{\perp} \dd m_{\perp}}\right)^{A-Z} .
\label{eq:coalymt}
\end{eqnarray}
%
Note that since the momenta and energies of protons and neutrons are equal, the proton and neutron distributions differ only on account of the different thermodynamic parameters used in equilibrium distribution functions. Note also that within our approximations $y_{\rm X}=y_{\rm p}=y_{\rm n}=y$, and at zero rapidity \refeq{eq:coalymt} reduces to the formula
\begin{eqnarray}
\left. \frac{\dd N_{\rm X}}{\dd y\, m_{\rm \perp X}^2 \dd m_{\rm\perp X}} 
\right|_{y=0} &=& \frac{A_{\rm FR}^{\rm X}}{\left(2\pi\right)^{A-1}}\left(
\left.\frac{\dd N_{\rm p}}{\dd y\, m_{\perp}^2 \dd m_{\perp}}
\right|_{y=0} \right)^Z\left(
\left.\frac{\dd N_{\rm n}}{\dd y\, m_{\perp}^2 \dd m_{\perp}} 
\right|_{y=0} \right)^{A-Z}.
\label{eq:coaly0}
\end{eqnarray}
Previous works introduce two models that reproduce well the spectra of protons and neutrons~\cite{Harabasz:2020sei,Harabasz:2022rdt}. Here we will use one of these models to predict, based on \refeq{eq:coaly0}, the spectra of $^3$H and $^3$He nuclei. For finite values of rapidity, \refeq{eq:coalymt} takes the form
\begin{eqnarray}
\frac{\dd N_{\rm X}}{\dd y\, m_{\rm \perp X}^2 \dd m_{\rm \perp X}} 
 &=& \frac{A_{\rm FR}^{\rm X}}{\left(2\pi \cosh y\right)^{A-1}}
\left( \frac{\dd N_{\rm p}}{\dd y\, m_{\perp }^2 \dd m_{\perp }}\right)^Z 
\left(\frac{\dd N_{\rm n}}{\dd y\, m_{\perp }^2 \dd m_{\perp }}\right)^{A-Z} .
\label{eq:coaly}
\end{eqnarray}
By integrating this equation over the transverse mass of the particle, we obtain the particle rapidity distribution.

\subsection{Nucleus formation rate}
\label{sec:Afr}
%
A popular form of the coefficient $A_{\rm FR}^{\rm X}$ used in the literature is~\cite{Mrowczynski:2016xqm}
\begin{equation}
A_{\rm FR}^{\rm X} = \frac{2s_{\rm X}+1}{\left(2s+1\right)^A} (2\pi)^{3\left(A-1\right)} \int\dd^3 r_1\dots\dd^3 r_A \,D(\rv_1,\dots,\rv_A) \, \left| \phi_{\rm X}(\rv_1,\dots,\rv_A)\right|^2,
\end{equation}
where the function $D(\rv_1,\dots,\rv_A)$ is the distribution of the space positions of neutrons and protons at freeze-out, normalized to unity, $\phi_X(\rv_1,\dots,\rv_A)$ is the light nucleus wave function, and $s$, $s_{\rm X}$ stand for the nucleon spin and spin of X, respectively. Many works on the coalescence model assume the Gaussian profile \cite{Mrowczynski:1992gc,
Mrowczynski:1987oid} 
\begin{equation}
D(\rv_1,\dots,\rv_A) = \left(4 \pi R_{\rm kin}^2 \right)^{-\f{3A}{2}}
\exp\left(- \frac{\sum_{i=1}^A \rv_i^2 }{4 R_{\rm kin}^2} \right),
\label{eq:DrG}
\end{equation}
where $R_{\rm kin}$ is the radius of the system at freeze-out. However, this formula may be considered inconsistent with the physical assumptions used in our model, where original particles are produced independently inside a~sphere of radius $R$ at a fixed laboratory time $t$~\cite{Harabasz:2020sei,Harabasz:2022rdt}. The expression (\ref{eq:DrG}) gives the root-mean-squared value $r_{\rm rms} = \sqrt{6} R \approx 2.45 R$, which implies hadron production far away from the original thermal system and its long formation time. Thus, as an alternative to the Gaussian distribution (\ref{eq:DrG}), we use the position distribution of particles that are produced independently within a~sphere of radius $R$ 
\begin{equation}
D(\rv_1,\dots,\rv_A) =\left(\frac{3}{4\pi R^3}\right)^A 
\theta_{\rm H}\left(R-|\rv_1|\right)\dots
\theta_{\rm H}\left(R-|\rv_A|\right),
\end{equation}
with the normalization condition
\begin{equation}
\int\dd^3r_1 \,\dd^3r_2 \, ... \,\dd^3r_A D(\rv_1,\dots,\rv_A) =1.
\end{equation}
Since in this work we are interested in the production of $^3$H and $^3$He nuclei, in the following we consider the case $A = 3$ and use the form
\begin{equation}
D(\rv_1,\rv_2,\rv_3) 
=\left(\frac{3}{4\pi R^3}\right)^3 \,
\theta_{\rm H}\left(R-|\rv_1|\right)
\theta_{\rm H}\left(R-|\rv_2|\right)
\theta_{\rm H}\left(R-|\rv_3|\right).
\end{equation}
Here $\theta_{\rm H}(x)$ is the Heaviside step function. In this case, it is convenient to introduce a set of normalized Jacobi coordinates
\begin{align}\begin{split}\label{eq:Jc}
\rv_{\rm cm} &= \frac{1}{3}\left(\rv_1+\rv_2+\rv_3\right), \\
\rv_{12} &= \rv_1-\rv_2,  \\
\rv_{312} \equiv \deltav &= \rv_3-\frac{1}{2}\left(\rv_1+\rv_2\right).
\end{split}\end{align}
The coordinates $\rv_{\rm cm}$, $\rv_{12}$, and $\rv_{312}$ can be interpreted as the center of mass of the X system, the relative distance between two particles, and the distance between this pair's center of mass and a third particle. Changing to these variables, we write
\begin{align}\begin{split}\label{eq:DRJ}
D(\rv_{\rm cm},\rv_{12},\rv_{312}) &= \left(\frac{3}{4\pi R^3}\right)^3  
\theta_{\rm H}\left(R-\left|\rv_{\rm cm}
+\frac{\rv_{12}}{2}-\frac{\rv_{312}}{3}\right|\right) \\
& \times \theta_{\rm H}\left(R-\left|\rv_{\rm cm}-\frac{\rv_{12}}{2}-\frac{\rv_{312}}{3}\right|\right)\theta_{\rm H}\left(R-\left|\rv_{\rm cm}+\frac{2 \,\rv_{312}}{3}\right|\right),
\end{split}\end{align}
\begin{figure}[t!]
\includegraphics[width=.7\linewidth]{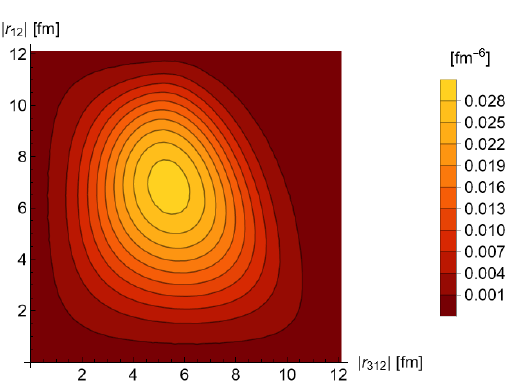}
\vspace{-0.5cm}
\caption{Nucleon space distribution integrated over the center of mass coordinate in terms of the moduli of the $\rv_{12}$ and $\rv_{312}$ coordinates, $ D(r_{12},r_{312}) = \left(4\pi\right)^2  r_{12}^2 r_{312}^2 \int \dd r^3_{\rm cm} D(r_{\rm cm},r_{12},r_{312})$.} \label{fig:figurephi1}
\end{figure}
\begin{figure}[t!]\includegraphics[width=0.49\textwidth]{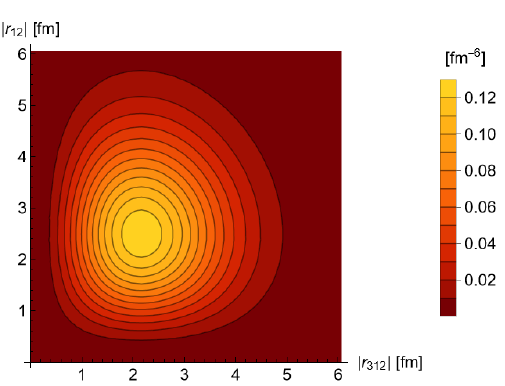} 
\includegraphics[width=0.49\textwidth]{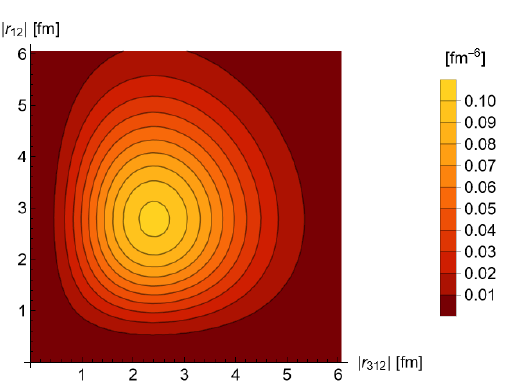}  
\vspace{-0.5cm}
\caption{Wave function space distribution in terms of the moduli of the $\rv_{12}$ and $\rv_{312}$ coordinates $\left(4\pi\right)^2 r_{12}^2 r_{312}^2 
\left| \phi_X(r_{12},r_{312})\right|^2$ in the Jacobi coordinates of $^3$H (left) and $^3$He  (right).}
\label{fig:figurephi2wf}
\end{figure}

For the $^3$H and $^3$He wave function, we use the Gaussian approximation in the same coordinates \cite{Bellini:2020cbj}
\begin{equation}
\left| \phi_X(r_{12},r_{312})\right|^2 = \left(3 \pi^2 R_{\rm X}^4 \right)^{-3/2}
\exp\left(- \frac{r_{12}^2+\frac{4}{3} r_{312}^2}{ 2R_{\rm X}
^2} \right),
\end{equation}
where $R_{\rm \,^3\!H} = 1.76~$fm  and $R_{\rm \,^3\!He} = 1.96~$fm are the radii of $^3$H and $^3$He nuclei, respectively. The function $\left| \phi_{\rm X}(r_{12},r_{312})\right|^2$ is normalized to unity. In Ref.~\cite{Florkowski:2023uim} it was argued that relativistic corrections can be neglected, and we continue this approach in this article.
We note that the wave function's dependence on its arguments can be factorized, allowing us to write
\begin{align}\begin{split}
A_{\rm FR}^{\rm X} &= \f{3\sqrt{3}}{4 R_{\rm X}^6 R^9}   \int \dd^3r_{\rm cm}\dd^3r_{12}  \tH \left(R-\left|\bo{r}_{\rm cm} + \f{\bo{r}_{12}}{2}\right|\right)\tH\left(R-\left|\bo{r}_{\rm cm} - \f{\bo{r}_{12}}{2}\right|\right) \exp \left( - \f{\bo{r}_{12}^2}{2 R_{\rm X}^2} \right)\\
&\times \int \dd^3r_{312} \tH\left(R-\left|\bo{r}_{\rm cm} + \bo{r}_{312}\right|\right) \exp \left( - \f{2\bo{r}_{312}^2 }{3 R_{\rm X}^2} \right).
\end{split}\end{align}
The inner integral is invariant under rotation of $\bo{r}_{\rm cm}$, so we can set
\begin{equation}
\bo{r}_{\rm cm} = (0,0,r_{\rm cm})
\end{equation}
and change the variables to
\begin{align}\begin{split}
\bo{r}_{\rm cm} + \bo{r}_{312} &= R \bo{\rho}_\delta \equiv R (x_\delta, y_\delta, z_\delta),\\
\bo{r}_{\rm cm} &= R \bo{\rho}_{c},\\
\bo{r}_{12} &= R \bo{\rho}_{12},
\end{split}\end{align}
which gives
\begin{align}\begin{split}
A_{\rm FR}^{\rm X} &= \f{3\sqrt{3}}{4 R_{\rm X}^6}   \int \dd^3\rho_{c}\dd^3\rho_{12}  \tH\left(1-\left|\bo{\rho}_{c} + \f{\bo{\rho}_{12}}{2}\right|\right)\tH\left(1-\left|\bo{\rho}_{c} - \f{\bo{\rho}_{12}}{2}\right|\right) \exp \left( - \f{\bo{\rho}_{12}^2 R^2}{2 R_{\rm X}^2} \right)\\
&\times \int \dd^3\rho_{\delta} \tH\left(1-|\bo{\rho}_\delta|\right ) \exp \left[ - \f{2 R^2}{3 R_{\rm X}^2} \left( x_\delta^2 + y_\delta^2 + \left(z_\delta - |\bo{\rho}_c|\right)^2\right)\right].
\end{split}\end{align}
The inner integral is equal to
\begin{align}
I_1(\rho_c, \xi) = \f{3 \pi}{8 \xi^4 \rho_c} \left[-3 \exp(-\phi^2) +3 \exp(-\chi^2) + \sqrt{6 \pi} \xi \rho_c (\rm{erf} \ \phi + \rm{erf} \ \chi) \right],
\end{align}
with
\begin{equation}
\xi = \f{R}{R_{\rm X}}, \quad \phi = \sqrt{\f{2}{3}} \xi (1- \rho_c), \quad \chi = \sqrt{\f{2}{3}} \xi (1+ \rho_c).
\end{equation}
In the outer integral, we change the variables again to get rid of the step functions
\begin{align}\begin{split}
\bo{\rho}_c + \f{\bo{\rho}_{12}}{2} &= \bo{a},\\
\bo{\rho}_c - \f{\bo{\rho}_{12}}{2} &= \bo{b},
\end{split}\end{align}
\begin{align}\begin{split}
A_{\rm FR}^{\rm X} &= \f{3\sqrt{3}}{4 R_{\rm X}^6}  \underset{\bo{a}, \bo{b} \in B(1)}{\int} \dd^3 a \, \dd^3 b  \exp \left[- \f{(\bo{b} - \bo{a})^2 \xi^2}{2} \right] I_1 \left(\f{|\bo{a}+\bo{b}|}{2},\xi \right),
\end{split}\end{align}
where $B(1)$ denotes a ball of radius 1 centered at the origin. The values of the formation rate coefficient for different freeze-out scenarios are shown in Table~\ref{tab:As}. The $^3$H and $^3$He wave functions are shown in Figs.~\ref{fig:figurephi1} and~\ref{fig:figurephi2wf}.

\begin{table}[t]
    \centering
    \begingroup
    \setlength{\tabcolsep}{13pt} 
    \renewcommand{\arraystretch}{1.2} 
    \begin{tabular}{|c|c|}
    \hline
    \hline
    Parameter & Value [MeV$^6$] \\
    \hline
    $A_{\rm FR}^{\rm ^3\!H}$  & $5.51\times 10^{11}$ \\
    $A_{\rm FR}^{\rm ^3\!He}$ & $5.03\times 10^{11}$ \\
    \hline
    \hline 
    \end{tabular}
    \endgroup
    \caption{Values of the formation rate parameter $A_{\rm FR}$ for $^3$H and $^3$He nuclei.
    }
    \label{tab:As}
\end{table}
%

\section{Freeze-out models}
\label{sec:fm}
%
The freeze-out models specify the hydrodynamic conditions for particle production at the latest stages of the system's spacetime evolution. In the single-freeze-out scenario~\cite{Broniowski:2001we,Broniowski:2001uk} adopted here, the freeze-out stage is defined by a set of thermodynamic variables such as temperature $T$, the baryon chemical potential $\mu_{\rm B}$, and the shape of the freeze-out hypersurface $\Sigma$. In addition, one defines the form of the hydrodynamic
flow $u^\mu(x)$ on $\Sigma$.

%
\subsection{Cooper-Frye formula}
%

The standard starting point for quantitative calculations is the Cooper-Frye formula \cite{Cooper:1974mv}, which describes the invariant momentum spectrum of particles
\begin{equation}
E \frac{\dd N}{\dd^3 p} = \int \dd^3\Sigma_\mu(x) \, p^\mu f(x,p).
\end{equation}
Here $f(x,p)$ is the phase-space distribution function of particles, and $p^\mu = \LR E, \pv \RR$ is their four-momentum with the mass-shell energy $E = \sqrt{m^2 + \pv^2}$.

The infinitesimal element of a three-dimensional freeze-out hypersurface from which particles are emitted $\dd^3\Sigma_\mu(x)$ may be obtained from the formula (see, for example, Ref.~\cite{Misner:1973prb})
\begin{equation}
\dd^3\Sigma_\mu = -\epsilon_{\mu \alpha \beta \gamma}
\frac{\partial x^\alpha}{\partial a }
\frac{\partial x^\beta }{ \partial b }
\frac{\partial x^\gamma }{ \partial c }\, \dd a\, \dd b\, \dd c,
\label{d3sigma}
\end{equation}
where $\epsilon_{\mu \alpha \beta \gamma}$ is the Levi-Civita tensor with the convention $\epsilon_{0123}=-1$, and $a, b, c$ are the three independent coordinates introduced to parametrize the hypersurface. This allows us to construct a six-dimensional, Lorentz-invariant density of the produced particles
\begin{equation}
   \dd^{6}N    =   \frac{\dd^3 p}{E} \,\,\dd^3\Sigma \cdot p \,\, f(x,p).
\end{equation}
The independent variables in such a general parametrization are the three components of three-momentum and the variables $a$, $b$, and $c$.

%
\subsection{Local equilibrium distributions}
%
%
In local equilibrium, the distribution function $f(x,p)$ has the following general form
\begin{equation}
f(x,p)= \frac{g_s}{(2\pi)^3}\LS \Upsilon^{-1} \exp\LR\frac{p \cdot u}{T}\RR-\chi \RS^{-1},
\end{equation}
where $T$ is the freeze-out temperature, $\chi=-1$ ($\chi=+1$) for Fermi-Dirac (Bose-Einstein) statistics, and $g_s = 2s +1$ is the degeneracy factor connected with spin. In this work we use such distributions of protons and neutrons only (as input for the light nuclei coalescence formula), hence we take $\chi=-1$ and $s=1/2$. Since we use the same mass for protons and neutrons, their equilibrium distributions differ only by the values of thermodynamic potentials that determine the fugacity factor
\begin{align}\begin{split}
\Upsilon_{\rm p} &=  \exp \left( \frac{\mu_{\rm B} + \frac{1}{2} \mu_{\rm I_3}}{T}\right), \\
\Upsilon_{\rm n} &=  \exp \left( \frac{\mu_{\rm B} - \frac{1}{2} \mu_{\rm I_3}}{T}\right).
\end{split}\end{align}
Here $\mu_{\rm B}$ and $\mu_{\rm I_3}$ are baryon and isospin chemical potentials, respectively. The values of $T$, $\mu_{\rm B}$, and $\mu_{\rm I_3}$ for the considered expansions scenario are given in Table~\ref{tab:params}. 

\section{Spheroidal expansion}
\label{sec:spd}
%
\subsection{Symmetry implementation}
%
%
In the case of freeze-outs that are spheroidally symmetric with respect to the beam axis, it is expedient to use the following parametrization of the spacetime points on the freeze-out hypersurface 
\begin{equation}
x^\mu(\zeta) = \LR t(\zeta), r(\zeta) \sqrt{1-\epsilon}\,\ev_{r\perp}, r(\zeta) \sqrt{1+\epsilon}\,
\cos\theta  \RR.
\end{equation}
Here $\ev_{r\perp} = \LR  \cos\phi\sin\theta,  \sin\phi \sin\theta\RR$, and the parameter $\epsilon$ controls deformation from a spherical shape. For $\epsilon~>~0$ the hypersurface is stretched in the (beam) $z$-direction.  The resulting infinitesimal element of the spheroidally symmetric hypersurface has the form 
\begin{widetext}
\begin{eqnarray}
\! \dd^3\Sigma_\mu \!&=& (1-\epsilon)\!\LR 
r^\prime  \sqrt{1+\epsilon}, 
t^\prime \frac{\sqrt{1+\epsilon}}{\sqrt{1-\epsilon}}\,\ev_{r\perp},
t^\prime \,\cos\theta \RR
r^2  \sin\theta \,\dd\theta\,\dd\phi\,\dd\zeta,\nn\\
\end{eqnarray}
\end{widetext}
where the prime denotes the derivatives taken with respect to $\zeta$. We also introduce the spheroidally symmetric hydrodynamic flow
\begin{eqnarray}
u^\mu &=& \gamma(\zeta,\theta)\LR 1,
v(\zeta) \sqrt{1-\delta}\,\ev_{r\perp}, 
v(\zeta) \sqrt{1+\delta}\,\cos\theta \RR, 
\end{eqnarray}
where $\gamma(\zeta,\theta)$ is the Lorentz factor, which is given by the formula 
\begin{equation}
\gamma(\zeta,\theta) = \LS 1-(1+\delta \cos(2 \theta))v^2(\zeta)\RS^{-1/2},
\end{equation}
resulting from the normalization condition $u\cdot u=1$.
The four-momentum of a hadron is parameterized as
\begin{equation}
p^\mu = \LR E , p_\perp\cos{\phi_p} , p_\perp\sin{\phi_p}, p_\parallel\RR,
\end{equation}
with $E=m_\perp \cosh(y)$, $p_\perp = \sqrt{m_\perp^2-m^2}$ and $p_\parallel=m_\perp \sinh(y)$. Thus, we obtain 
\begin{equation}
 u \cdot p=\gamma(\zeta,\theta) \left[ E\!-\, v(\zeta) \kappa (\delta) \right]
\label{puspheroid}
\end{equation}
and
\begin{equation}
\dd^3\Sigma \cdot p = \sqrt{1-\epsilon}\LR E \,r^\prime \sqrt{1-\epsilon^2} 
- t^\prime   \kappa(-\epsilon) \RR r^2 \sin\theta
\,\dd\theta \, \dd\phi \, \dd\zeta,
\end{equation}
where $\kappa (\xi)=   \sqrt{1+\xi}\, p_\parallel\cos\theta + \sqrt{1-\xi}\, p_\perp\sin\theta \cos(\phi-\phi_p)$.

Prior analyses of the spectra indicated that a satisfactory description of the data can be obtained by assuming $t^\prime=0$ (which allows putting $\zeta = r$), $\epsilon=0$, $\delta \neq 0$. In this case,
\begin{equation}
\dd^3\Sigma \cdot p =  E \, r^2  \dd r \sin\theta
\,\dd\theta \, \dd\phi ,
\end{equation}
and the Cooper-Frye formula for fermions takes the form
\begin{eqnarray}
\frac{dN}{dy \, m_\perp^2 dm_\perp} &=& \cosh y \,\, {\tilde S}(p,\theta_p),
\label{eq:spd1}
\end{eqnarray}
where 
\begin{eqnarray}
{\tilde S}(p,\theta_p)\!&=\!&\frac{g_s }{(2\pi)^2} \int\limits_0^R dr \, r^2 \int\limits_0^\pi d\theta \sin\theta  \int\limits_0^{2\pi} d\phi  \left[\Upsilon^{-1} \exp\left( \frac{ u\cdot p }{T} \right) + 1 \right]^{-1}. \label{eq:tS} 
\end{eqnarray}
The scalar product $u\cdot p$ is given by \refeq{puspheroid}, and spherical symmetry allows us to put $\phi_p = 0$. Changing the variables from $p$ and $\theta_p$ to rapidity and transverse mass, we write
\begin{eqnarray}
\frac{dN}{dy \, m_\perp^2 dm_\perp} &=& \cosh y \,\, 
{\tilde S}\left[ \sqrt{m_\perp^2 \cosh^2 y -m^2}, \theta_y(m_\perp,y) \right],
\label{eq:spd2}
\end{eqnarray}
where
\begin{equation}
\theta_y(m_\perp, y)  = \arccos 
\frac{m_\perp \sinh y}{\sqrt{m_\perp^2 \cosh^2 y -m^2}}.
\label{eq:thetay}
\end{equation}
\begin{figure*}[t]
\includegraphics[width=0.7\textwidth]{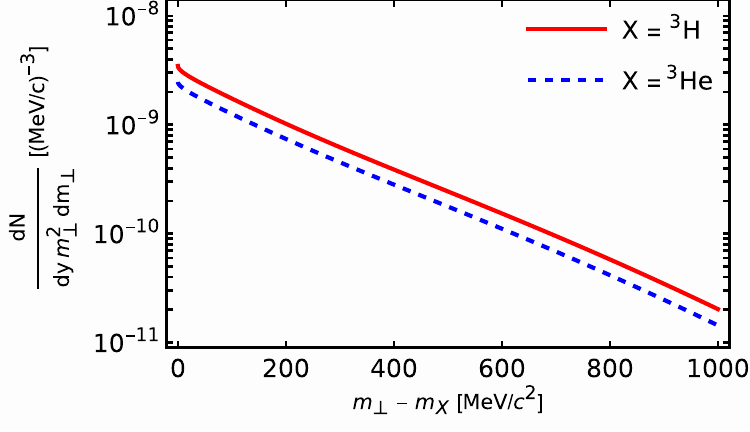} 
\vspace{-0.5cm}
\caption{Predicted transverse-mass spectra at zero rapidity for $^3$H (red solid line) and $^3$He (blue dashed line).}
\label{fig:spectra_t}
\end{figure*}

\begin{figure*}[t]
\includegraphics[width=0.7\textwidth]{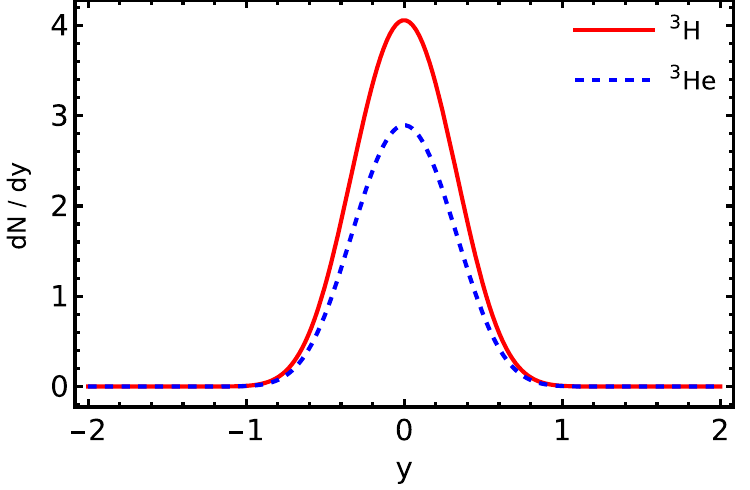}
\vspace{-0.5cm}
\caption{Predicted rapidity spectra of $^3$H (red solid line) and $^3$He (blue dashed line).}
\label{fig:spectra_y}
\end{figure*}

Within our approximations, the angle (\ref{eq:thetay}) is the same for nucleons and nuclei. In the zero-rapidity case,
\begin{eqnarray}
\left. \frac{dN}{dy \, m_\perp^2 dm_\perp} \right|_{y=0 }&=&  
{\tilde S}\left(p_\perp, \frac{\pi}{2} \right).
\label{eq:spd3}
\end{eqnarray}

\section{RESULTS}
\label{sec:results}

The model calculations of the spectra of $^3$H ($A=3, Z=1$) and $^3$He ($A=3, Z=2$) nuclei are based on~\refeq{eq:coalymt} with the nucleon distributions defined by Eq.~(\ref{eq:spd2}), where for protons and neutrons we use different values of the fugacity factor $\Upsilon$. Our results for the rapidity distributions and transverse-mass spectra are shown in Figs.~\ref{fig:spectra_t} and \ref{fig:spectra_y}.

Integrating the spectra, we find the total yields, which we give and compare to the experimental results in Table~\ref{tab:Nd}. We observe that the model calculations underestimate the experimental results by factors of 2.7 and 2.0 (for $^3$H and $^3$He, respectively). The main uncertainty of the theoretical results comes from the formation rate factors that are calculated in a relatively simple way described in Sec.~\ref{sec:Afr}. More realistic calculations of the formation rate may improve the agreement with the data (here we have in mind, for example, more realistic forms of the wave functions). We note that the values of the coefficients $A_{\rm FR}$ are very large (see Table~\ref{tab:As}), which means that our final results are products of very large and small numbers (all in units of MeV). Therefore, it is quite intriguing that we correctly reproduce the order of magnitude of the yields. This is achieved by the assumption that the coalescence takes place just after freeze-out within a sphere of the radius $R$. We note that other theoretical frameworks, for example, IQMD + MST, also underestimate the yields; see Ref.~\cite{IQMD,IQMDOrigin}.

\begin{table}[t]
    \centering
    \begingroup
    \setlength{\tabcolsep}{13pt}
    \renewcommand{\arraystretch}{1.2} 
    \begin{tabular}{|c|c|c|c|}
    \hline
    Nucleus & Quantity & Model & Experiment \cite{Szala:2019,Szala2019a}  \\
    \hline
    $^3$H & $N$             & 3.16\comment{3.164} & $8.65 \pm (0.01)_{\rm stat} \pm (1.05)_{\rm sys}$   \\
    & $(\dd N/\dd y)_{y=0}$ & 4.06\comment{4.058} & –   \\
     \hline
    $^3$He & $N$  & 2.26\comment{2.256} & $4.55 \pm (0.01)_{\rm stat} \pm (0.29)_{\rm sys}$   \\
    & $(\dd N/\dd y)_{y=0}$   & 2.89\comment{2.894} & –   \\
     \hline
    \end{tabular}
    \endgroup
    \caption{Model predictions for the total yield $N$ and particle yield at zero rapidity $(\dd N/\dd y)_{y=0}$  compared to the experimental results~\cite{Szala:2019,Szala2019a}.
    }
    \label{tab:Nd}
\end{table}

Additional verification of the model can be obtained from a comparison of the rapidity and transverse-momentum spectra. In fact, very often only the correct description of the shapes of the spectra of composite particles is regarded as the indication that the coalescence mechanism is at work. To check whether we can reproduce the rapidity profiles of the preliminary HADES data~\cite{Szala:2019,Szala2019a}, we have varied the values of $A_{\rm FR}$ by rescaling them by a factor $\alpha$ (compared to the original values derived above). In the case of the rapidity distribution of $^3$H, the best description is obtained for $\alpha = 2.3$, which is slightly smaller than the factor 2.7 needed to reproduce the experimental yield. For $^3$He, the best adjustment is obtained for $\alpha = 1.7$, compared to the factor 2.0 found above to fit the yield, see Fig.~5.~\footnote{The fit is performed by the minimization of the quantity $Q^2(\alpha) = (1/N) \sum_{i=1}^N (\alpha R_{i, \rm{model} }-R_{i, \rm{exp}} )^2/ R_{i, \rm{exp}}^2$, where $R_{i, \rm{model}}$ and $R_{i, \rm{exp}}$ are the model and experimental results, respectively.} The corresponding total multiplicities are given in Table \ref{tab:Nd2}, where we also show the dependence on the integration range in rapidity. 

On account of the uncertainty of the normalization of the experimental transverse-mass spectra, we have refrained from making comparisons between theory and experiment in this case. However, we observe that our model slopes of the $m_{\perp}$-distributions are steeper than those shown as preliminary results. 

\section{Summary and conclusions}
\label{sec:summary}

In this work, we have used the previously constructed model of hadron production in heavy-ion collisions in the few-GeV energy regime~\cite{Harabasz:2020sei,Harabasz:2022rdt,Florkowski:2023uim} to determine the yields and spectra of $^3$H and $^3$He. Our approach combines the thermal framework with single freeze-out with an idea of the coalescence of emitted nucleons. This approach turned out to be very successful in determining the yield of $^2$H \cite{Florkowski:2023uim}. The present work extends the previous framework to the next lightest nuclei. Our predictions for the yields of $^3$H and $^3$He are smaller by factors of 2.7 and 2.0 than the experimental values. Despite this discrepancy, it is noteworthy that our simple model correctly describes the order of magnitude of the multiplicities studied. We also note that the model yield values are controlled by the formation rate factors $A_{\rm FR}$ whose calculation can be improved in the future by using more realistic wave functions of $^3$H and $^3$He. The comparison of the preliminary HADES results on the spectra shows that we are missing a factor of about 2 to reproduce the rapidity spectra.

\begin{figure}[t]\includegraphics[width=\textwidth]{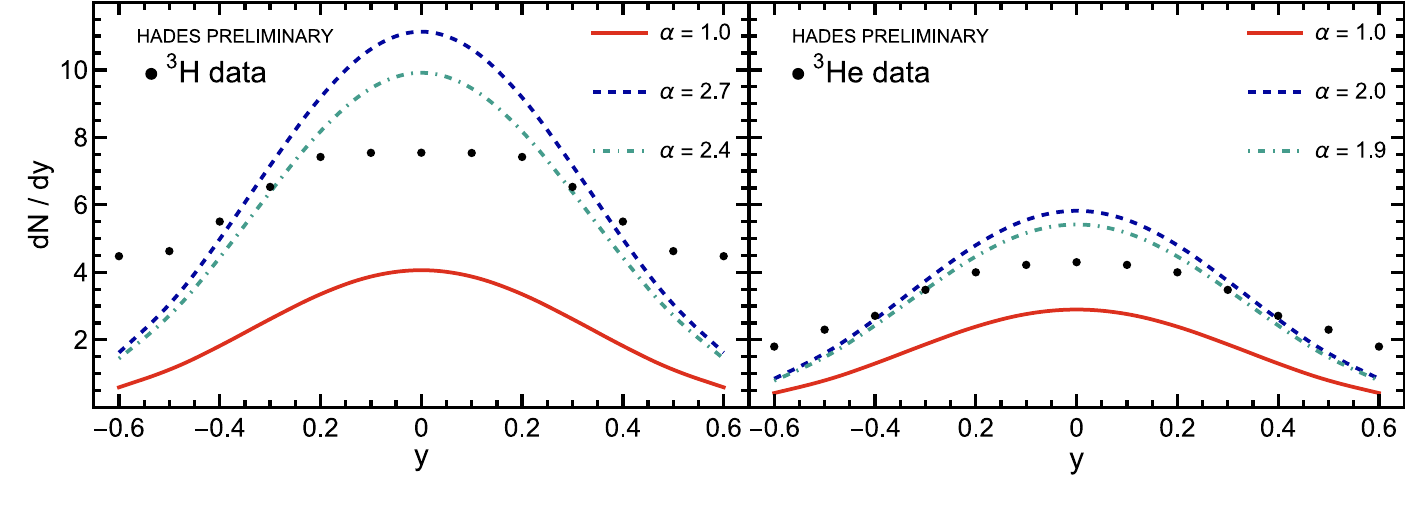} 
\vspace{-1.5cm}
\caption{Rapidity spectra prediction for $^3$H (left) and $^3$He (right) for different scaling factors: exact model value (red solid line, $\alpha =1.0$), fit to the experimental yield (blue dashed line, $\alpha =2.7$ for $^3$H and  $\alpha =2.0$ for $^3$He), and best fit to data (teal dash-dotted line, $\alpha =2.4$ for $^3$H and  $\alpha =1.9$ for $^3$He). The data shown are preliminary HADES results~\cite{Szala:2019,Szala2019a}.}
\label{fig:RapidityScaled}
\end{figure}

\begin{table}[h]
    \centering
    \begingroup
    \setlength{\tabcolsep}{13pt}
    \renewcommand{\arraystretch}{1}
    \begin{tabular}{|c|c|c|c|c|}
    \hline
    Nucleus & Scaling factor $\alpha$ & $\int^{0.65}_{-0.65}$ & $\int^{\infty}_{-\infty}$& Experiment \cite{Szala:2019,Szala2019a}  \\
    \hline
    $^3$H & $1.0$             & 3.10\comment{3.103}  & 3.16\comment{3.164} & $8.65 \pm (0.01)_{\rm stat} \pm (1.05)_{\rm sys}$   \\
    & $2.7$ & 8.50\comment{8.502} & 8.67\comment{8.671} & –   \\
    & $2.4$ & 7.02\comment{7.017} & 7.16\comment{7.156} & –   \\
     \hline
    $^3$He & $1.0$  & 2.21\comment{2.207} & 2.26\comment{2.256} & $4.55 \pm (0.01)_{\rm stat} \pm (0.29)_{\rm sys}$   \\
    & $2.0$   & 4.44\comment{4.437} & 4.55\comment{4.549} & –   \\
    & $1.9$   & 3.75\comment{3.755} & 3.85\comment{3.846} & –   \\
     \hline
    \end{tabular}
    \endgroup
    \caption{Model predictions for the total yield $N$ calculated for different scaling factors $\alpha$ and different integration ranges compared to the experimental results~\cite{Szala:2019,Szala2019a}.
    }
    \label{tab:Nd2}
\end{table}

\begin{acknowledgments} We thank Małgorzata Gumberidze, Szymon Harabasz, and Piotr Salabura for useful discussions concerning the experimental data. This work was supported in part by the Polish National Science Centre (NCN) Grants No.~2022/47/B/ST2/01372 (W.F.) and No.~2018/30/E/ST2/00432 (R.R.).
\end{acknowledgments}

\providecommand{\noopsort}[1]{}\providecommand{\singleletter}[1]{#1}%

\end{document}